\documentclass{aa}  
\usepackage{graphicx}
\usepackage{txfonts}

\usepackage{supertabular}
\usepackage{natbib}
\usepackage[flushleft]{threeparttable}
\def\kms{\,$\mathrm{km\, s^{-1}}$}

\newcommand{\logg}{\ensuremath{\log g}}

\begin{document}
%
\title{Follow-up observations of extremely metal-poor stars identified from SDSS. \thanks{Based on observations obtained with the Hobby-Eberly Telescope, which is a joint project of the University of Texas at Austin, the Pennsylvania State University, Stanford University, Ludwig-Maximilians-Universit\"at M\"unchen, and Georg-August-Universit\"at G\"ottingen.}}

   \author{D.~S. Aguado\inst{1,2}, C. Allende Prieto\inst{1,2}, J.~I . Gonz\'alez Hern\'andez\inst{1,2}, R. Carrera\inst{1,2}, 
   R. Rebolo\inst{1,2,3}, M. Shetrone\inst{4}, D. L. Lambert\inst{4},  E. Fern\'andez-Alvar\inst{5} }

   \institute{Instituto de Astrof\'{\i}sica de Canarias,
              V\'{\i}a L\'actea, 38205 La Laguna, Tenerife, Spain\\              
         \and
             Universidad de La Laguna, Departamento de Astrof\'{\i}sica, 
             38206 La Laguna, Tenerife, Spain \\  
         \and
             Consejo Superior de Investigaciones Cient\'{\i}ficas, 28006 Madrid, Spain\\
         \and McDonald Observatory and Department of Astronomy, University of Texas, Austin, TX 78712, USA\\
            \and Instituto de Astronom\'ia, Universidad Nacional Aut\'onoma de M\'exico, AP 70-264, 04510 Ciudad de M\'exico, M\'exico
            }

%
\authorrunning{D.S. Aguado et al.}
\titlerunning{EMP stars identified from SDSS}

\date{Received March 15, 2016; accepted June 1, 2016}

 
  \abstract
   {The most metal-poor stars in the Milky Way witnessed the early phases of formation of the Galaxy, 
   and have chemical compositions that are close to the pristine mixture from Big Bang 
   nucleosynthesis, polluted by one or few supernovae.}
   { Only two dozen stars with ([Fe/H]$< -4$) are known, and they show a wide range of abundance 
 patterns. It is therefore important to enlarge this sample.
 We present the first results of an effort to identify new extremely metal-poor stars in the Milky Way halo.}  
   {Our targets have been selected from low-resolution spectra obtained as part of the 
   Sloan Digital Sky Survey, and followed-up with medium resolution spectroscopy on the 
   4.2 \,m William Herschel Telescope and, in a few cases, at high resolution on the the 9.2 \,m 
   Hobby-Eberly Telescope.
   Stellar parameters and the abundances of magnesium, calcium, iron, and strontium 
   have been inferred from 
   the spectra using classical model atmospheres. We have also derived carbon abundances from 
   the G band.}
   {We find consistency between the metallicities estimated from SDSS and those from new data at the level of 0.3\,dex.
   The  analysis of medium resolution data obtained with ISIS on the WHT allow us to refine
   the metallicities and in some cases measure other elemental abundances.
       Our sample contains 11 new metal-poor stars with
   $\left[{\rm Fe/H}\right]<-3.0$, one of them with an estimated metallicity
   of $\left[{\rm Fe/H}\right]\sim-4.0$. We also discuss metallicity discrepancies of some stars in common with previous works in the literature. 
   Only one of these stars is found to be C-enhanced at about
   [C/Fe]$ \sim +1$, whereas the other metal-poor stars show C abundances at the
   level of [C/Fe]$\sim +0.45$.}


   \maketitle
%

\section{Introduction\label{Intro}}

The oldest stars in the Milky Way belong to the halo and thick-disk populations (see, e.g. \citet{red06,hay13}.
The majority of the halo stars we can date seem to be older than about 10 \,Gyr,
while stars in the thick disk can be as young as $\sim$ 8 \,Gyr \citep{alle06}.
The metallicity distribution of the halo is broad (FWHM$\sim1.2$\,dex), with
an average metallicity of  
about $\left[{\rm Fe/H}\right]<-1.5$, which has been 
recently found to shift to lower values at distances from the Galactic center of
about 30 kpc \citep{car08,yong12,chen14,alle14,emm15}. 

Compared to the disk population, the stellar halo has a very low density (about 1\% of 
all stars in the solar neighborhood) but it becomes the dominant population 
at distances from the plane larger than 4-5 \,kpc. Because these halo stars are far, 
they are difficult to observe. Deep spectroscopic surveys have been very helpful to study
this population and, for example, 
over one million halo stars have spectroscopy from the Sloan Digital Sky
Survey (SDSS,\citealt{yan09,eis11}).

The oldest halo stars are the most interesting, particularly those that formed in the first 
or second generation \citep{norris13I,kel14}, and which must have primitive 
compositions, in particular very low
metal abundances. Such objects are extremely rare as illustrates the fact that, despite 
substantial efforts, only 4 stars are known at [Fe/H]$<-5$. 
Theoretical calculations show that the lack of metals 
in the gas in the early universe prevents gas clouds from fragmenting effectively, and 
shifts the initial mass function to very high masses. The implication is that no low-mass stars 
were formed in the first generation \citep{bro03, boni11, boni12}. However, the minimum metallicity at 
which low-mass stars can form appears lower than suggested by theoretical predictions, 
and needs to be assessed 
empirically, as at least one unevolved F-type star,  SDSS J10291+1729 \citep{caff11}, 
not only shows [Fe/H]$\simeq -5$, but also low C and O abundances.
Furthermore, it has been suggested that the C/Fe abundance ratios in extremely 
metal-poor stars exhibit a bimodal distribution \citep{caff14,alle15}, which may be the 
result of two main types of supernovae coexisting in the early phases of evolution 
of our Galaxy.

The fraction of stars with large carbon enhancements increases
significantly for low iron abundances \citep{cohen09,boni15}. These so-called Carbon enhanced metal-poor stars (CEMP) 
constitute about 30\%
of stars at [Fe/H]$<-3.0$, 40\% at [Fe/H]$<-3.5$, and 75\% at [Fe/H]$<-4$. Moreover,
the four stars known at [Fe/H]$<-5$  fall in this category. Currently, J10291+1729
sets the metallicity limit for the formation of low-mass stars, but is this an
anomaly? are there many other stars with even lower iron abundances which are not
enhanced in carbon or oxygen?

\begin{figure}
\begin{center}
{\includegraphics[angle =180,width=90 mm]{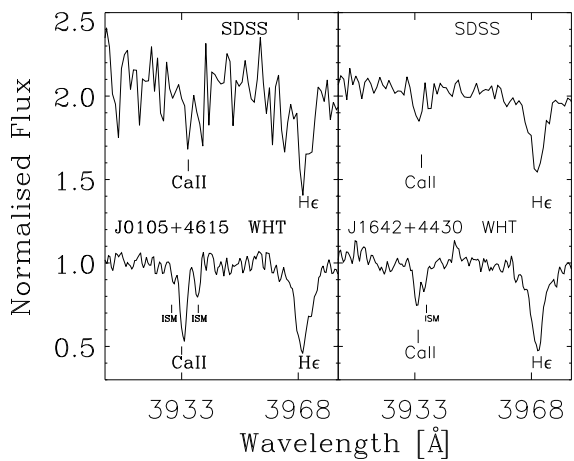}}
\end{center}
\caption{Medium resolution SDSS spectra ($R\sim1800$, top spectra) of two extremely 
metal-poor candidates, J0105+4615 (left panel) and J1642+4430 (right panel), together with 
ISIS spectra  ($R\sim2400$, bottom spectra) obtained at the 4.2m-WHT telescope. 
The \ion{Ca}{ii} K spectral lines from the star and the ISM are identified.
}
\label{ism}
\end{figure}

Solving these issues requires larger samples of extremely metal-poor stars (EMP).
We present here a sample selected from SDSS spectra 
and observed on
the William Herschel Telescope in La Palma, equipped with the medium resolution 
spectrograph ISIS.  In Fig. \ref{ism} we show an example of the quality of the SDSS spectra and the ISIS spectra in the 
CaII H\&K spectral region.
Two of the stars have also been observed at higher resolution using HRS on the
Hobby-Eberly Telescope at McDonald Observatory. Section \ref{TargetSelection} 
describes how the candidates were identified from SDSS spectra, \S \ref{Observations} 
provides an account of the observations carried out and the data reduction, \S \ref{Analysis} 
details our spectroscopic analysis, and \S \ref{Conclusions} summarizes our results and 
conclusions.

\section{SDSS analysis and target selection}
\label{TargetSelection}

For a sample of more than a million objects with spectra consistent with zero redshift from the Sloan 
Extension for Galactic Understanding and Exploration (SEGUE,
\citealt{yan09}) and/or the Baryonic Oscillations Spectroscopic Survey (BOSS, \citealt{eis11,daw13}),
we derive the stellar parameters: effective temperature $T_{\rm eff}$, surface gravity 
$\log g$, and metallicity [Fe/H] \footnote{We use the bracket notation for reporting
chemical abundances: [a/b]$ = \log \left( \frac{\rm N(a)}{\rm N(b)}\right) - \log
\left( \frac{\rm N(a)}{\rm N(b)}\right)_{\odot}$,
where $\rm N$(x) represents number density of nuclei of the element x.}.

The SDSS optical spectra are from SDSS Data Release 9 (DR9, \citealt{dr9})
for observations with the original SDSS spectrograph,  
and DR12 \citep{dr12} for data obtained with the upgraded BOSS spectrographs \citep{smee13}. 
These optical spectra have modest signal-to-noise ratios and a resolving power of about 2,000.  
The spectral range 3850 -- 9190\,\AA 
is matched to model spectra computed with classical model atmospheres. 

We use an automatic pipeline based on   
{\tt FER\reflectbox{R}E}\footnote{{\tt FER\reflectbox{R}E} is available 
from http://hebe.as.utexas.edu/ferre} \citep[]{alle06}.
The code divided the spectra in about 300/400$\sim$\AA pieces, which are normalized to
their mean fluxes, effectively removing low-frequency systematic errors in flux
calibration and ISM extinction. The model spectra used in the analysis are treated
exactly in the same fashion. 
The grid of synthetic spectra is the one described by \citet{alle14} 
 but with the addition of the C-enhanced models.
The carbon abundance is derived as a free parameter in the range 
$-1.0 <$~[C/Fe]~$ < +4.0$, whereas the $\alpha$-element is set to [$\alpha$/Fe]=$+0.4$.

This analysis determines simultaneously the  main three stellar
parameters and the carbon abundances, assuming a micro-turbulence of 2 km s$^{-1}$.
Table~\ref{basic} provides the derived atmospheric parameters and carbon abundances, as well as the 
magnitudes, equatorial coordinates, and signal-to-noise ratio for each SDSS spectrum. 

\begin{table*}
\begin{center}
 \renewcommand{\tabcolsep}{5pt}
\centering

\caption{Coordinates and atmospheric parameters for the program 
stars based in the analysis of the SDSS spectra with the FERRE code. The FERRE internal uncertainties
values in this table, discussed in section \ref{TargetSelection} are shown in brackets.
\label{basic}}
\begin{tabular}{lccccccccccc}
\hline
Star & $g$ &      RA &      DEC &$T_{\rm eff}$ & $\log g$  & $\left[{\rm Fe/H}\right]$ 
& $\left[{\rm C/Fe}\right]$ &$<S/N>^{a}$\\
     & mag & J2000 & J2000 & K & $cm\, s^{-2}$ &               &                       \\
     &     & h  '   ''&$\mathring{}$  '  ''  &   &                 &       &                           \\
\hline\hline
SDSS J010505+461521    &19.3  & 01:05:05.88  &+46:15:21.60&5480 &1.79& $ -3.86 $&-0.48 &15  \\
SDSS J014036+234458    &15.8  & 01:40:36.22  &+23:44:58.20&6092 &4.77& $ -3.46 $& 1.01 & 55\\
SDSS J021958$-$084955  &16.4  & 02:19:58.25  &-08:49:55.92&5705 &3.97& $ -3.31 $& 0.35 &  61\\
SDSS J040114$-$051259  &18.6  & 04:01:14.71  &-05:12:59.06&5854 &4.98& $ -3.82 $&-0.79 &  39\\
SDSS J044655+113741    &18.6  & 04:46:55.70  &+11:37:41.16&5970 &2.89& $ -3.30 $& 0.67 & 23\\
SDSS J063055+255243    &18.1  & 06:30:55.57  &+25:52:43.72&5411 &1.08& $ -3.61 $& 0.78&  42\\
SDSS J075818+653906    &17.7  & 07:58:18.28  &+65:39:06.95&6078 &3.90& $ -3.25 $& 0.71 &  25\\ 
SDSS J101600+172901    &16.6  & 10:16:00.43  &+17:29:01.32&5416 &4.92& $ -3.40 $& 0.01 &  49\\
SDSS J120441+120111    &16.4  & 12:04:41.38  &+12:01:11.64&5852 &3.91& $ -3.55 $&-0.50 & 61\\
SDSS J123055+000547    &14.8  & 12:30:55.25  &+00:05:47.04&6199 &4.18& $ -3.30 $&-0.25 &  51\\ 
SDSS J132250+012343    &16.3  & 13:22:50.59  &+01:23:43.08&5247 &1.11& $ -3.58 $& 0.78 &  62\\
SDSS J164234+443004    &17.8  & 16:42:34.48  &+44:30:04.96&6282 &4.98& $ -4.61 $& 0.55 & 32\\ 
SDSS J165618+342523    &15.7  & 16:56:18.31  &+34:25:23.16&5014 &1.23& $ -3.08 $&-0.47 & 65\\ 
SDSS J183455+421328    &19.1  & 18:34:55.03  &+42:13:28.92&5367 &4.92& $ -3.94 $&-0.45 & 17\\ 
SDSS J214633$-$003910  &18.1  & 21:46:33.17  &-00:39:10.08&6469&4.90& $ -4.85 $& 0.57 & 18\\ 
SDSS J220646$-$092545  &15.4  & 22:06:46.20  &-09:25:45.84&4983 &1.03& $ -3.07 $& 0.45 & 53\\
\hline
\hline
\end{tabular}
\end{center}
\begin{tablenotes}
 \item $^a$ Signal-no-noise ratio have been calculated as average of the SDSS entire spectrum
 
\end{tablenotes}

\end{table*}

\begin{figure*}
\begin{center}
{\includegraphics[angle =180,width=180 mm]{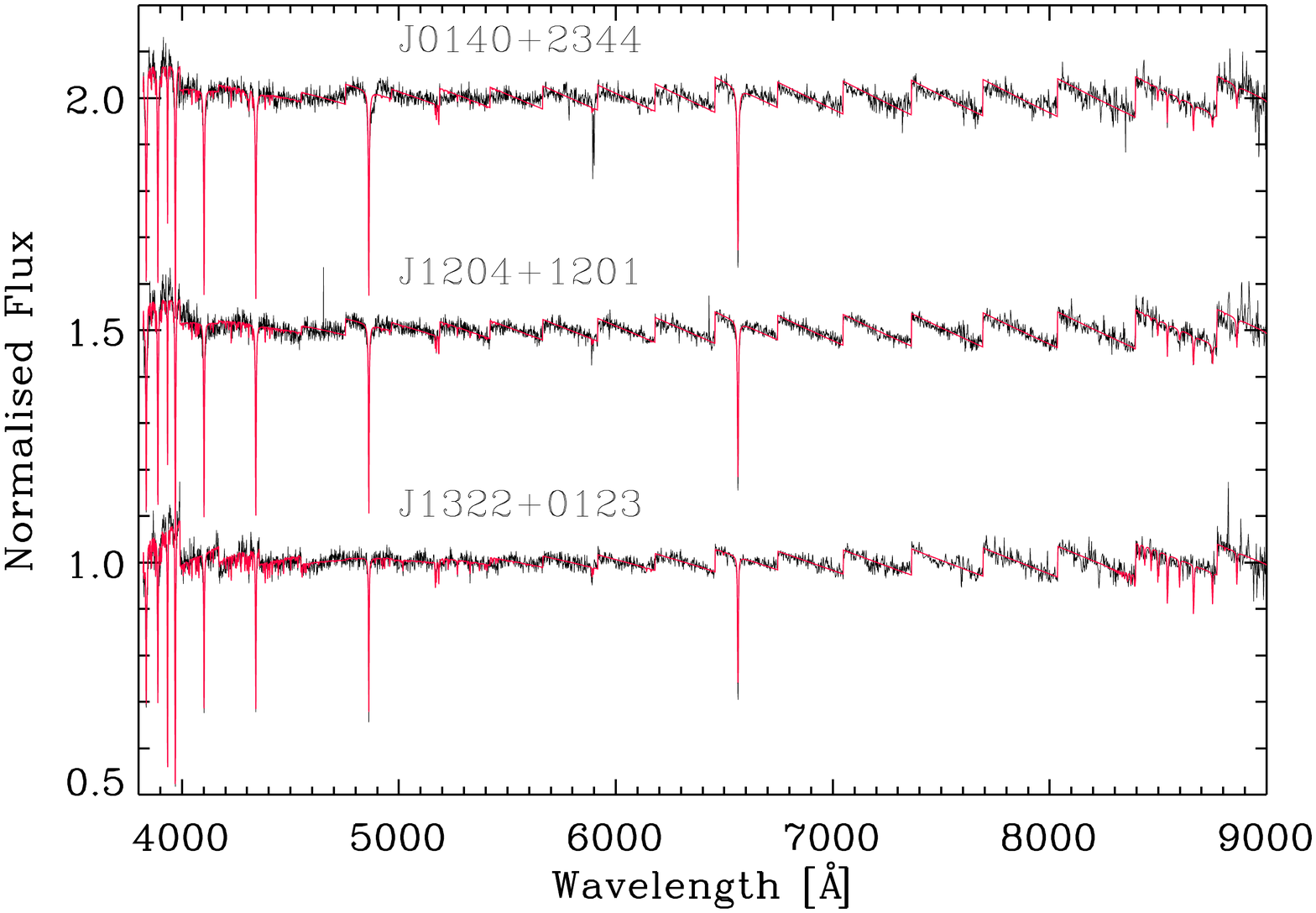}}
\end{center}
\caption{SDSS spectra of J0140+2344, J1204+1201 and J1322+0123 (black line) anaysed with.
A reliable flux calibration allows to derive effective temperature using the slope of the continuum (red line).}
\label{compared}
\end{figure*}

\begin{figure}
\thispagestyle{empty}
\begin{center}
{\includegraphics[width=90 mm]{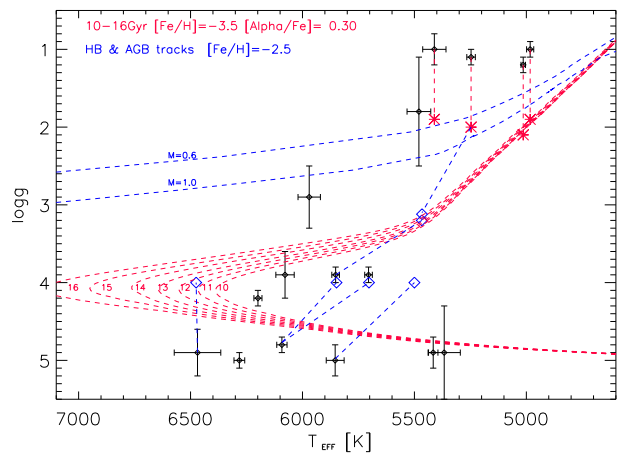}}
\end{center}
\caption{
DARMOUTH isochrones for $\left[{\rm Fe/H}\right]=-3.5$ and differents ages from 16 to 10 Gyr (red dashed lines), 
 blue dashed lines are HB and AGB theoretical tracks for $\left[{\rm Fe/H}\right]=-2.5$ 
 for two different relative masses (M=0.6 and M=1.0). The black diamonds represent the stars of this 
 work and its internal uncertainties derived from FERRE analysis.
 The red crosses are the four low-gravity objects explained in Section \ref{TargetSelection}.
 The blue diamonds are bibliography values from
 \citet{yong13II, caff13I, pla15} and discussed in Section \ref{compa}.}
\label{isochrone}
\end{figure}

The effective temperatures and surface gravities derived from  SDSS spectra 
were, in most cases, adopted for the subsequent analysis of the Spectrograph and Imaging System (ISIS) spectra.
 We consider our determination of effective temperatures very reliable.
These are based on the fitting of the whole SDSS spectral range and 
of the local stellar continuum and Balmer lines, in particular (see Fig. \ref{compared}).
The estimated uncertainties have been discussed
in detail by \citet{alle14}, who compared the results from {\tt FER\reflectbox{R}E} with those from 
the SEGUE Stellar Parameter Pipeline \citep[SSPP,][]{lee08II, lee08I}, finding
a rms scatter between the two sets of values for halo stars of 
about $\sigma_{T_{\rm eff}}=70$ \,K,  $\sigma_{logg}=0.24$ and  $\sigma_{\left[{\rm Fe/H}\right]}=0.11$.
However, systematic errors for low gravity ($\log g<3$) stars have been detected from the comparison with
the SSPP and the analysis of the globular
cluster M13 \citep{alle14}. 
Taking those offsets into account the preferred gravity for J010505+461521 is $\log g=1.9$, 
for J063055+255243 is is $\log g=1.9$, 
for J132250+012343 is $\log g=2.0$, and for  J165618+342523 is $\log g=2.1$.
Since our stars are on the low-metallicity edge
of the distribution of halo stars, our uncertainties are likely somewhat
higher than those inferred from these statistics.
 Fig.~\ref{isochrone} shows DARMOUTH isochrones \footnote{The Dartmouth Stellar Evolution Program (DSEP) is available
from www.stellar.dartmouth.edu},  HB and AGB tracks compared to the stellar parameters derived
with {\tt FER\reflectbox{R}E} and its uncertainties.
For this paper we adopt the following errors 
$\sigma_{T_{\rm eff}}=150$ \,K, $\sigma_{logg}=0.3$ and $\sigma_{\left[{\rm Fe/H}\right]}=0.2$\,dex
(see also Section \ref{wht_w2044}.

We selected about a hundred EMP candidates from SDSS spectroscopy.
All  selected candidates have been classified in three priority levels, 
according to the quality of the spectrum, the goodness-of-fit for the Balmer lines,
and the $g$-band magnitude. The signal-to-noise ratio varies across the sample, and
we selected spectra with signal-to-noise ratios S/N$>15$, with emphasis in targets at S/N$>30$. The brightness distribution of the candidates makes it 
very difficult to find EMP candidates with high signal-to-noise ratios (exposition time is constant).

We included in our observations a few 
 well-known metal poor stars, which are very useful to  compare the performance of
our methods with the results from the literature. More details are provided in the
following section.

\section{Observations and data reduction}
\label{Observations}

\subsection{Observations with ISIS on the 4.2m WHT}

\begin{table*}
\renewcommand{\tabcolsep}{1pt}
\centering
\begin{center}
\begin{tabular}{lcrccc}

\end{tabular}
\end{center}
\end{table*}

We obtained long-slit spectroscopy with ISIS \citep{isiswht}, attached
to the 4.2-m William Herschel Telescope (WHT), at the Roque
de los Muchachos Observatory (La Palma) over the course of 
four observing runs; Run I: August 17-19 (3 nights), 2012; II: March 23-25 (3 nights), 2013; III: Dec 31 - Jan 2 (3 nights), 2015; IV: February 5-8 (4 nights), 2015.
Twenty nine objects were observed in total. 
We used the R600B and R600R gratings,  with the default dichroic, and a 
GG495 filter was used on the red arm. The observations were made with a one-arcsecond wide slit, the 
resolving power was $R\sim2400$ at 4500\,\AA  in the blue arm, and $R\sim5400$
in the red arm.
The instrument configuration was the standard ISIS R600B/R600R and the
spectral ranges covered by the blue and red ISIS arms are 3500-5200\,\AA
and 7420-9180\,\AA, respectively.

The information on the exposures obtained for each target is given in Table \ref{observations}.
 We lost about for nights of a total of 13 awarded for this project because of bad weather conditions.
For purpose, we selected two targets 
identified from the HAMBURG-ESO survey, HE\,1327-2326 \citep{fre05}, 
one of the most metal-poor stars known, and HE\,1523-0901\citep{fre07},
a strongly r-process-enhanced VMP in whose spectrum uranium has been detected. In addition,
2MASS J\,2045-2842, a well-known ultra metal-poor star with moderately low effective temperature, 
$T_{\rm eff}\simeq 4750$ K, was observed.

Data reduction included bias substraction, flat-fielding, wavelength 
calibration (using CuNe $+$ CuAr lamps), 
and combination of individual spectra, was performed using the \emph{onespec} package in IRAF 
\footnote{IRAF is distributed by the National Optical Astronomy Observatory, 
which is operated by the Association of Universities for Research in Astronomy 
(AURA) under cooperative agreement with the National Science Foundation} \citep{tod93}.

\begin{figure*}
\begin{center}
{\includegraphics[width=180 mm]{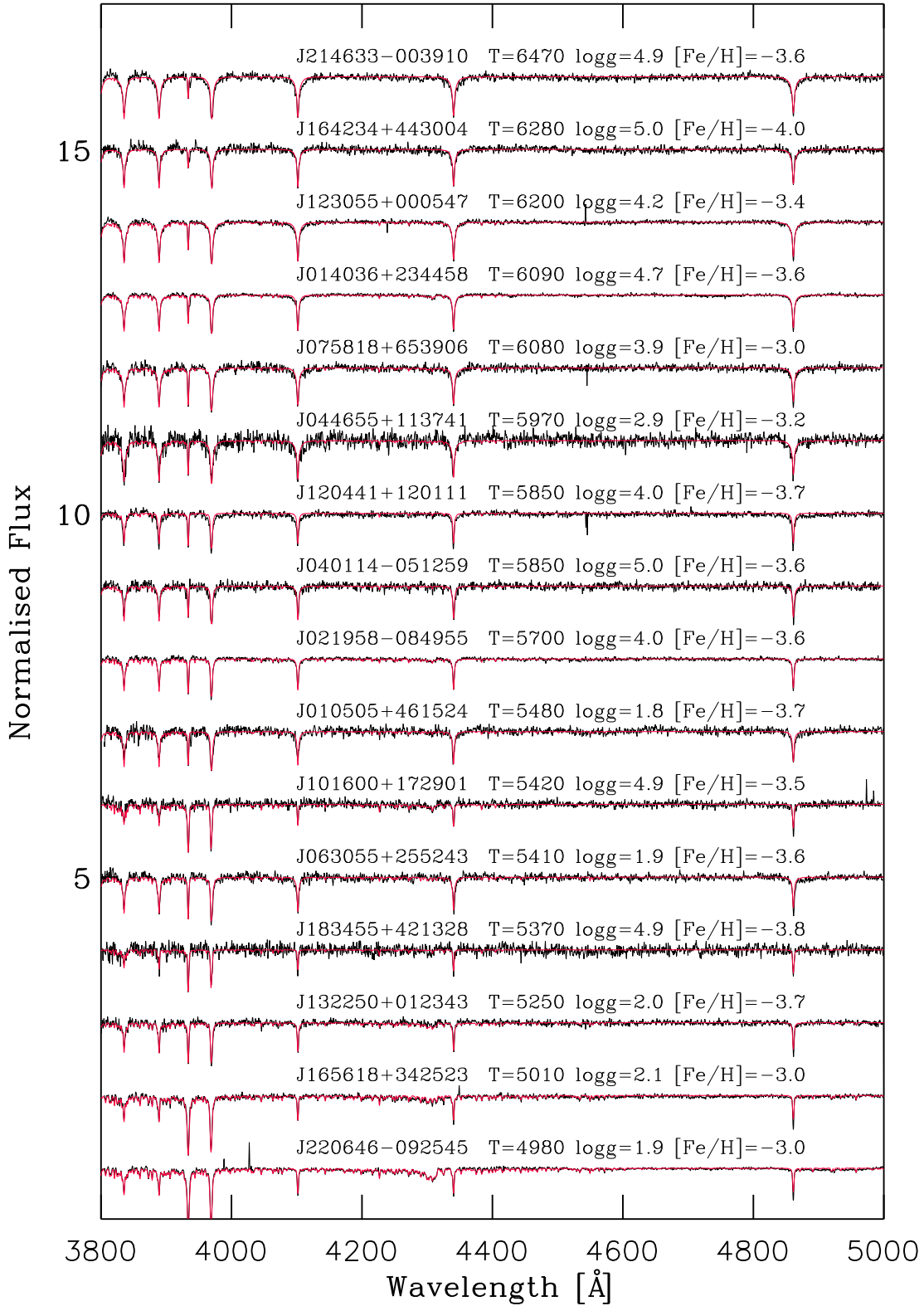}}
\end{center}
\caption{The ISIS/WHT blue arm spectra (3800\,\AA-5000\,\AA) from the full sample
(black line) and the best fit calculated with SYNTHE (red line). Over each spectrum the main stellar parameters
are plotted.}
\label{cal}
\end{figure*}

The ISIS spectra, with a resolution somewhat higher that those from SDSS,  improve
our chances of resolving
potential contributions from the Interstellar Medium (ISM) to the absorption 
in the vicinity of the \ion{Ca}{ii} K line. 
In Fig.~\ref{ism} we show several examples corresponding to candidates in which at least 
part of the ISM contribution can be identified in the ISIS spectrum while the SDSS low resolution do not allow us to do it. 
However, there are cases in which the ISM and stellar absorption cannot be resolved 
and our calcium abundances (and the corresponding iron abundances inferred from them) 
are, strictly speaking, upper limits.
 In Fig. \ref{cal} we depict the ISIS spectra together with synthetic spectra computed with the SYNTHE 
code (see Section \ref{wht_analysis}) of all stars of our sample.

\begin{figure*}
\begin{center}
{\includegraphics[width=190 mm]{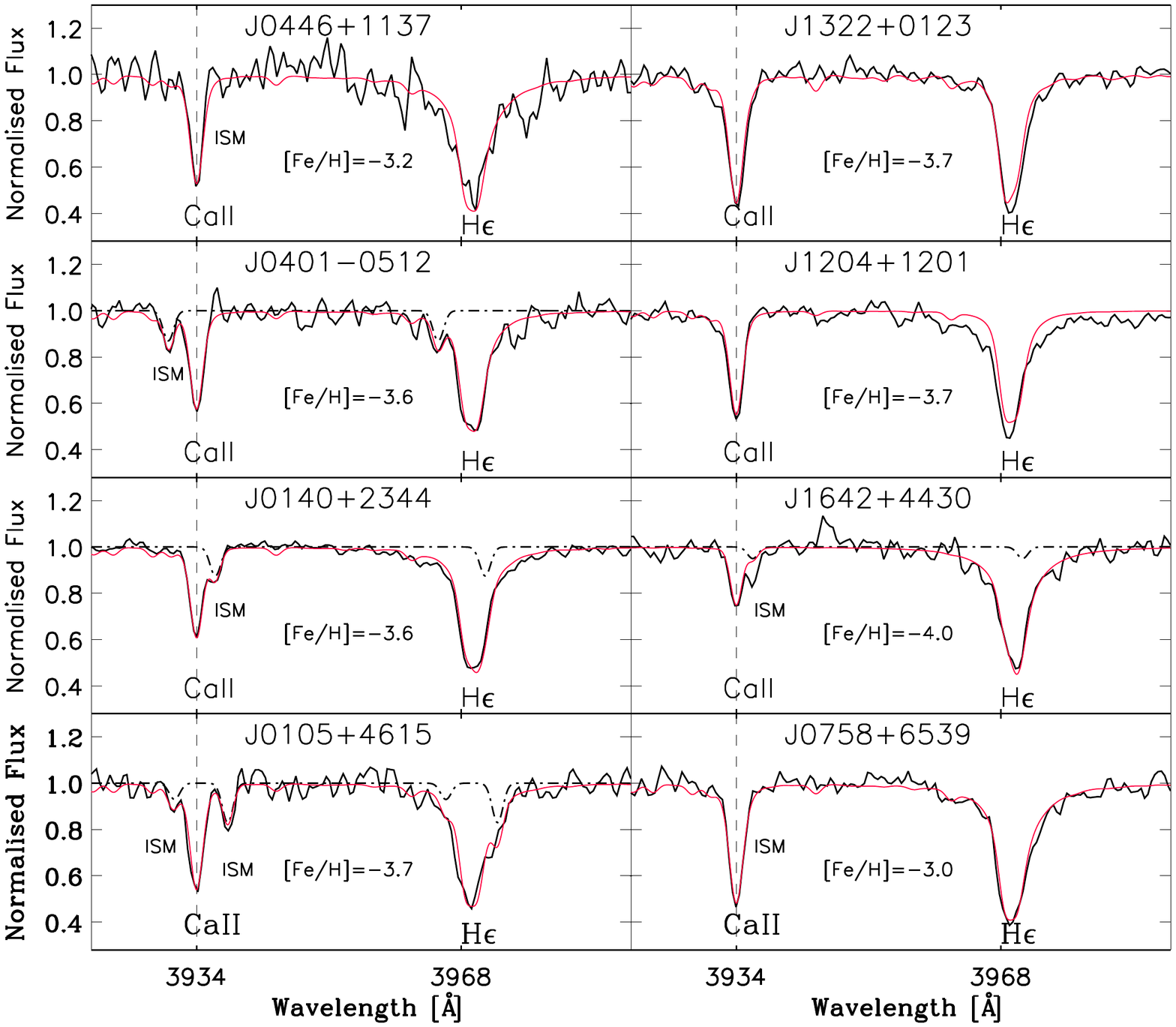}}
\end{center}
\caption{ISIS spectra (solid black line) together with best fit synthetic spectra 
(solid red line) in the \ion{Ca}{ii} region for six different EMP,
J0401$-$0512 (upper-left) with $\left[{\rm Fe/H}\right]=-3.6$,
J0140+2344 (middle-left) with $\left[{\rm Fe/H}\right]=-3.6$,
J0105+4615 (lower-left) with $\left[{\rm Fe/H}\right]=-3.7$, 
J1642+4430 (upper-right) with $\left[{\rm Fe/H}\right]=-4.0$,
J0758+6539 (middle-right) with $\left[{\rm Fe/H}\right]=-3.0$,
J0446+1137 (lower-right) with $\left[{\rm Fe/H}\right]=-3.2$,
after the ISIS/WHT analysis. 
The ISM contribution in calcium K and H lines are added to the synthetic 
spectra (dashed-black line) calculated with SYNTHE.\label{cal_comp_ump}}
\end{figure*}

\subsection{Observations with HRS on the 9.2m HET}

The HRS \citep{tull98} observations on HET \citep{ram98} were obtained in service mode \citep{she07} over the period November 11, 2012 -- 
March 22, 2013 at the McDonald Observatory (Texas). A total of 15.5 hours were 
allocated to this program, of which 2.5 and 8\,hr were used for J0140+2344 and J0219$-$0849, respectively. 
The spectral range of these spectra spans 4000-5400\,\AA, with
a gap between 4700 and 4800\,\AA. The resolving power was $R\sim15000$, with 3.2 pixels
per resolution element. The HRS configuration included two sky fibers, and 
$2\times5$ binning on the CCDs.
HRS/HET frames were processed with IRAF task \emph{ccdproc} and \emph{apflatten}. 
The extraction and wavelength-calibration  were performed with the \emph{echelle} package tasks within IRAF.
The sky subtraction was performed with our own tools using the two HRS sky fibers.
Finally the echelle orders of each spectrum were merged and normalized using the \emph{norchelle} task developed
by one of us \footnote{Available from Allende Prieto web page http://www.as.utexas.edu/~hebe/stools/}.

\section{Analysis}
\label{Analysis}

\subsection{Analysis of ISIS spectra with MOOG and SYNTHE}\label{wht_analysis}

Using the atmospheric parameters from the SDSS analysis, 
we calculate custom model atmospheres using ATLAS9 acording to \citet{mez12}. 
We measure the equivalent widths of the \ion{Ca}{ii} K line with the \emph{splot} IRAF routine.
Then, we derive  calcium abundances analyzing the spectra with MOOG \citep{sne73}.
These values are given in Table \ref{AnalysisResults}.
Here we also assume $\left[{\rm \alpha/Fe}\right]$=0.4 and derive metallicities with $\left[{\rm Fe/H}\right]$=$\left[{\rm Ca/H}\right]$-$\left[{\rm \alpha/Fe}\right]$.
Then we check for consistency between the values obtained 
from the SDSS spectra and the calcium abundances from the equivalent-width method.
The atomic data adopted for the lines analysed in this work are listed 
in Table~\ref{lines}. 
 The solar abundance values adopted in this paper are A($\rm Mg$)$=7.53$, A(Ca)$=6.31$, 
A(Sr)$=2.92$ and A(Fe)$=7.45$ \citep{asp05}.

\begin{figure*}[]
\begin{center}
{\includegraphics[width=140 mm, angle =180]{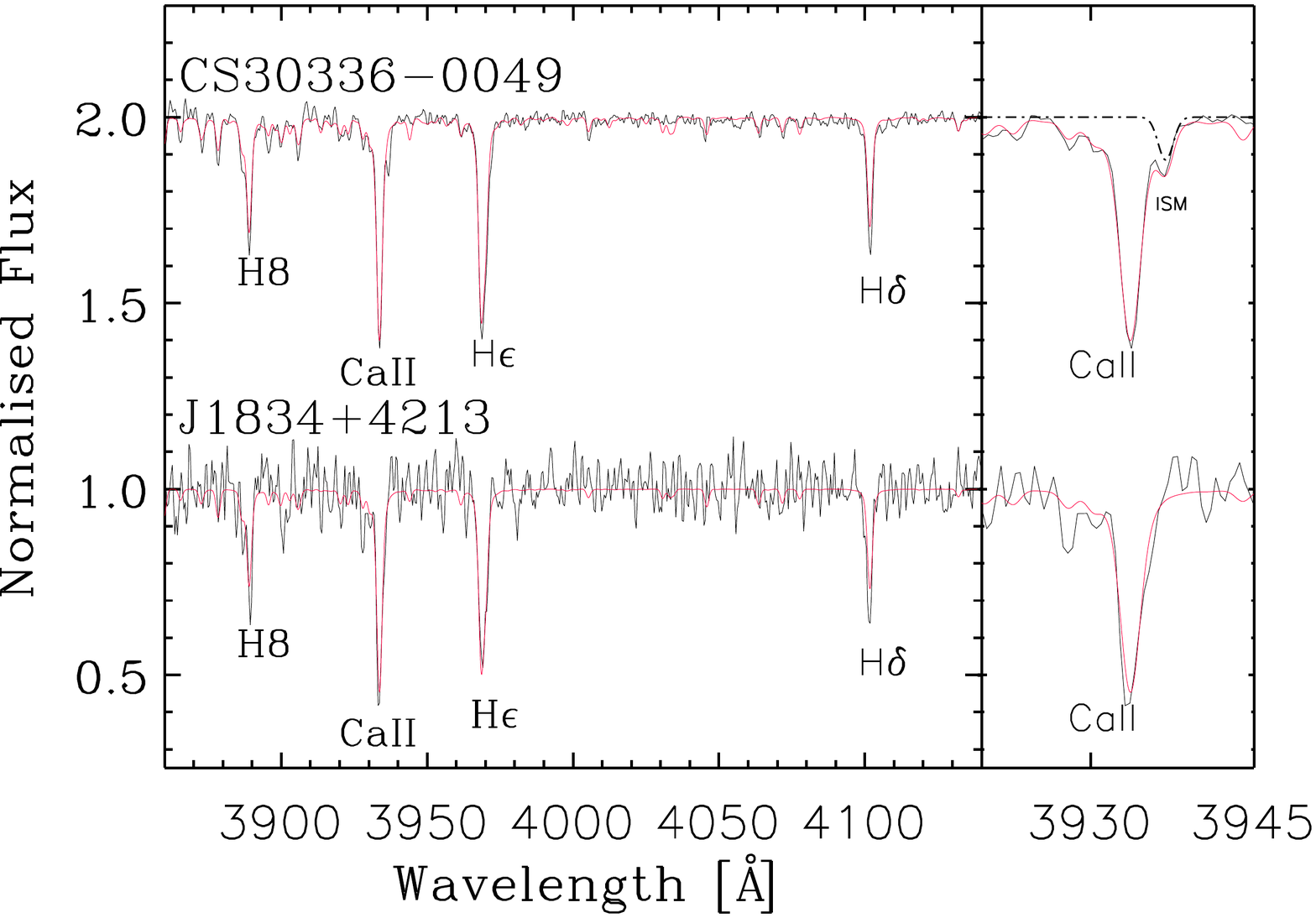}
\includegraphics[width=140 mm]{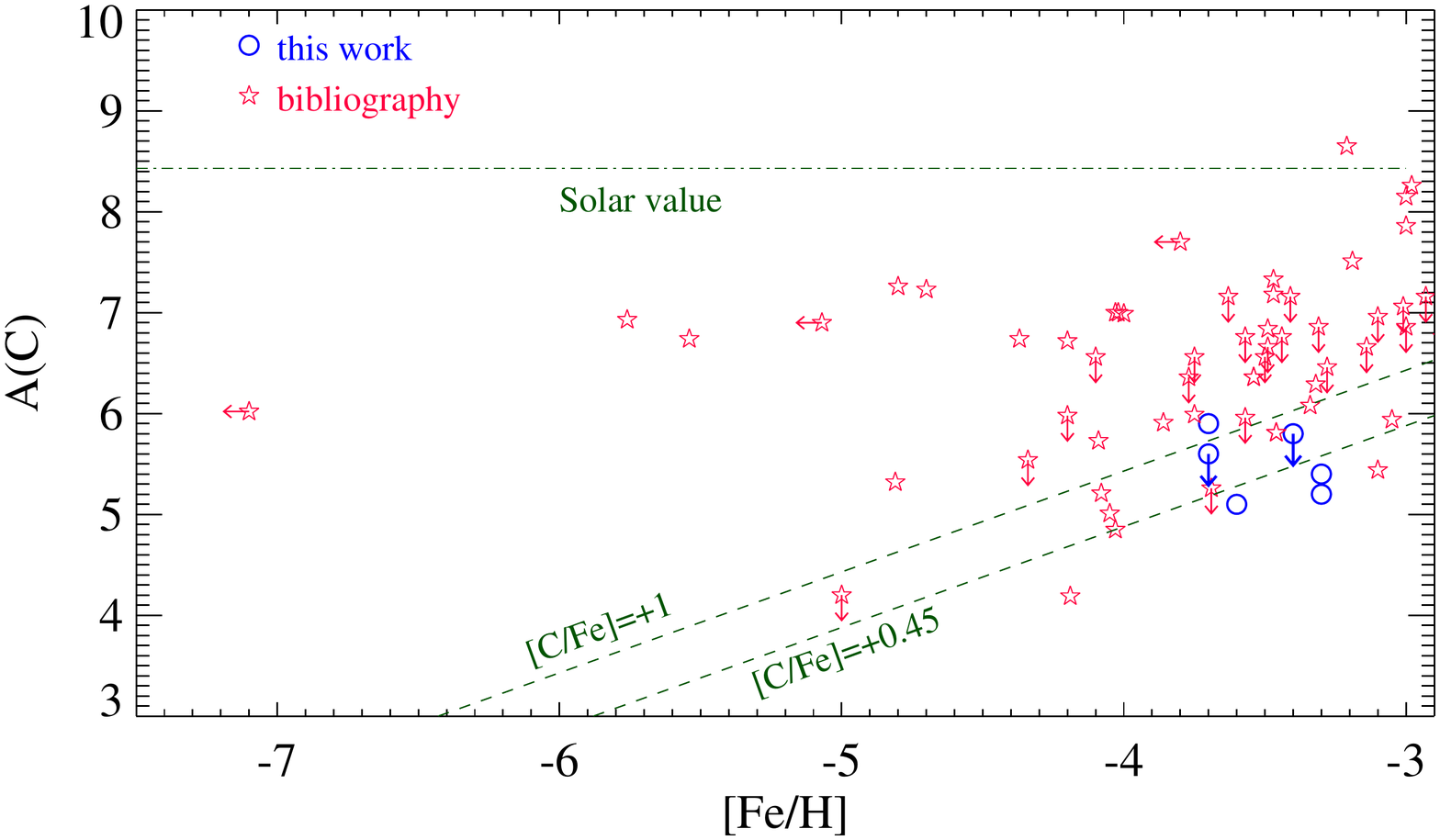}}
\end{center}
\caption{{\it Upper panel}: ISIS spectra of CS30336-0049 (upper), a well-known 
metal-poor star with $T_{\rm eff}=4750$, $\logg=1.19$ and $\left[{\rm Fe/H}\right]=-4.1$,
compared with SDSS J1834+4213 (lower), a new EMP with  $T_{\rm eff}=5370$, 
$\logg=4.9$ and $\left[{\rm Fe/H}\right]=-3.8$. 
 Carbon abundances versus metallicity of CEMP stars. Stars analysed in this work are represented by circles.  
The other stars (stars symbol) come from \citep{siv06,yong13II,fre05,fre06,caff14,alle15}. 
The  upper-dashed line represents the solar value A(C)=8.43 with very few CEMP stars over this value. 
A CEMP star defined as  $\left[{\rm C/Fe}\right]>+1$ is plotted dashed line \citep{bee05} and
the mean carbon-to-iron ratio of turnoff extremely metal-poor stars, defined as [C/Fe]$>+0.45$ 
\citep{boni09} is represented by dashed line.}
\label{cal_comp_ump2}
\end{figure*}

The second and more thorough analysis of the ISIS spectra 
was performed using the code SYNTHE \citep{kur05,sbo05}.
We adopt the effective temperatures and surface gravities from SDSS
spectra with the exception explained in section \ref{TargetSelection}.
Then manually optimize metallicity for each star.  
In nearly 85\% of the cases the metallicity value inferred in this fashion
is very close to the one from the analysis of the SDSS observations, 
suggesting that our SDSS results are fairly reliable.
 Table~\ref{AnalysisResults} summarizes the results from the SYNTHE analysis.
The average offset of both metallicity determinations is [Fe/H]$_{\rm SDSS}-$[Fe/H]$_{\rm WHT}=+0.21$ 
with a standard
deviation of $0.31$. 
Fig.~\ref{cal_comp_ump} illustrates the agreement between model and ISIS blue-arm
observations for eight of the stars in our sample in which there are several obvious 
ISM contributions to the absorption in the vicinity of the stellar
\ion{Ca}{ii} K line. The top panel of Fig.~\ref{cal_comp_ump2} shows a well-known metal poor star, 
 CS30336-0049 and one of our new identified metal-poor stars J1834$-$4213.

\begin{table*}
\begin{center}
\renewcommand{\tabcolsep}{5pt}
\centering
\caption{The stellar parameters and main results obtained from ISIS spectra.
\label{AnalysisResults}}
\begin{tabular}{lcccccccccccc}
\hline
Star              &   & ${\rm T_{\rm eff}}$ & \logg & $\xi$ &  $[\rm Fe/H]_{\rm SYNTHE}$  & EW$_{\rm CaII K}$&
A(Ca)$_{\rm MOOG}$ & A(Ca)$_{\rm SYNTHE}$&A(C)$_{\rm SYNTHE}$\\
                    &  & [K] & [cm\, s$^{-2}$] & [\kms] &  & [\AA] & \\

\hline\hline
SDSS\,J010505+461521 & &5480 & 1.9 & 2.0   & $-3.7\pm 0.2$ &$1.2\pm 0.1 $&$3.1\pm 0.4$ & $3.0\pm 0.2$ &  -  \\
SDSS\,J014036+234458 && 6090 & 4.7 & 2.0   & $-3.6\pm 0.2$ &$0.7\pm 0.1 $&$3.0\pm 0.4$ & $3.2\pm 0.2$ &  $5.9\pm 0.4$    \\
SDSS\,J021958$-$084955 && 5700 & 4.0 & 2.0   & $-3.6\pm 0.2$ &$1.3\pm 0.1$ &$3.2\pm 0.4$  & $3.2\pm 0.2$ & $<5.6$     \\
SDSS\,J040114$-$051259 && 5850 & 5.0 & 2.0   & $-3.6\pm 0.2$  &$ 1.0\pm 0.1$ &$3.0\pm 0.4$ & $3.2\pm 0.2$ &   -    \\
SDSS\,J044655+113741 & &5970 & 2.9 & 2.0   & $-3.2\pm 0.2$ & $0.9\pm 0.1$&$3.3\pm 0.4$  & $3.6\pm 0.2$ &  -    \\
SDSS\,J063055+255243 & &5410 & 1.9 & 2.0   & $-3.6\pm 0.2$ &$1.4\pm 0.1 $&$3.0\pm 0.4$  & $3.2\pm 0.2$ &  -     \\
SDSS\,J075818+653906 & &6080 & 3.9 & 2.0   & $-3.0\pm 0.2$ &$1.2 \pm 0.1$&$3.5\pm 0.4$  & $3.8\pm 0.2$ &   -     \\
SDSS\,J101600+172901 & &5420 & 4.9 & 2.0   & $-3.5\pm 0.2$ &$2.1\pm 0.1$&$3.1\pm 0.4$ & $3.3\pm 0.2$ &  $5.1\pm 0.2$&     \\
SDSS\,J120441+120111 & &5850 & 3.9 &2.0   & $-3.7\pm 0.2$  &$1.0\pm 0.1$ &$2.9\pm 0.4$  & $3.1\pm 0.2$ & - \\
SDSS\,J123055+000547 & &6200 & 4.2 &2.0   & $-3.4\pm 0.2$  &$0.9 \pm 0.1$&$3.3\pm 0.4$  & $3.4\pm 0.2$ & $<5.8$ \\
SDSS\,J132250+012343 & &5250 & 2.0 & 2.0   & $-3.7\pm 0.2$ & $1.7 \pm 0.1$&$3.2\pm 0.4$  & $3.1\pm 0.2$ &  -     \\
SDSS\,J164234+443004 && 6280 & 5.0 & 2.0   & $-4.0\pm 0.2$ &$0.5\pm 0.1$&$2.8\pm 0.4$  & $2.8\pm 0.2$ & - &     \\
SDSS\,J165618+342523 && 5010 & 2.1 & 2.0   & $-3.3\pm 0.2$ &$2.7\pm 0.1$&$3.3\pm 0.4$  & $3.5\pm 0.2$ & $5.2\pm 0.2$ &  \\
SDSS\,J183455+421328 & &5370 & 4.9 & 2.0   & $-3.8\pm 0.2$  & $1.5\pm 0.1$&$2.8\pm 0.4$  & $2.9\pm 0.2$ & -    \\
SDSS\,J214633$-$003910 && 6470 & 4.9 & 2.0   & $-3.6\pm 0.2$  & $0.7\pm 0.1$&$3.2\pm 0.4$  & $3.2\pm 0.2$ & -    \\
SDSS\,J220646$-$092545& & 4980 & 1.9 & 2.0  & $-3.0\pm 0.2$  &$3.2\pm 0.1 $&$3.5\pm 0.4$  & $3.8\pm 0.2$ & $5.4\pm 0.2$  \\
\hline

\end{tabular}
\end{center}
\begin{tablenotes}
\item $-$The T$_{\rm eff}$ has been adopted from the FERRE results (see Table \ref{basic}).
\item $-$The $\logg$ has been adopted from the FERRE results in way shows in \ref{TargetSelection}.
\item $-$The T$_{\rm eff}$ and $\logg$ uncertainties are described in \ref{TargetSelection}.
 \item $-$The $\left[{\rm Fe/H}\right]$ has been derived from the $\left[{\rm Ca/H}\right]$ assuming $\left[{\rm \alpha/Fe}\right]=0.4$,
 following the expression $\left[{\rm Fe/H}\right]$=$\left[{\rm Ca/H}\right]$-$\left[{\rm \alpha/Fe}\right]$
 \item $-$The $\left[{\rm C/Fe}\right]$ has been derived only in the cases the quality of the ISIS spectra is high enough.
\end{tablenotes}

\end{table*}

\begin{table}

\caption{
Atomic lines analysed in this work.\label{lines}}
\begin{center}
\begin{tabular}{lrrr}
\hline\noalign{\smallskip}
 Element & $\lambda$   & ${\rm E.P.}$ & log\,gf \\
         & $\mathring{A}$        & eV  & \\
\hline\noalign{\smallskip}
\ion{Mg}{i}  & 5167.321 & 2.709 & $-0.931$ \\  
\ion{Mg}{i}  & 5172.684 & 2.712 & $-0.450$ \\ 
\ion{Mg}{i}  & 5183.604 & 2.717 & $-0.239$ \\ 
\ion{Ca}{i}  & 4226.728 & 0.000 & $+0.265$ \\ 
\ion{Ca}{ii} & 3933.663 & 0.000 & $+0.135$ \\ 
\ion{Fe}{i}  & 5328.038 & 0.9146 & $-1.466$  \\ 
\ion{Fe}{ii}  & 5169.028 & -0.87 & $+2.891$  \\ 
\ion{Sr}{ii}  & 4077.714 & 0.000 & $+0.148$  \\ 
\ion{Sr}{ii}  & 4215.552 & 0.00 & $-0.173$  \\ 
\noalign{\smallskip}\hline\noalign{\smallskip}

\end{tabular}
\end{center}                   
\end{table}


The equivalent-width method allows us to derive calcium abundances.
The abundance uncertainty for this method is estimated by the expression:
\begin{equation}\label{eq}
\begin{split}
 \sigma^2_{[\rm Ca/H]}=\left(\frac{\partial[Ca/H]}{\partial T_{\rm eff}}\right)^2 \sigma^2_{T_{\rm eff}}+\left(\frac{\partial[Ca/H]}{\partial EW}\right)^2 \sigma^2_{EW}+\\
 +\mathcal{O}\left(\logg\right)+\mathcal{O}\left(\left[{\rm Fe/H}\right]\right)
 \end{split}
\end{equation}

 neglecting the terms depending on  $\logg$ and $\left[{\rm Fe/H}\right]$ 
We estimate an uncertainty of 0.1 \AA in the EW$_{\rm CaII K}$ values measured with \emph{splot}.
The uncertainties in Teff are given in Section \ref{TargetSelection}.
Following this procedure we obtain $0.3$\,dex$<\sigma_{[\rm Ca/H]}<0.4$\,dex for the Ca abundances derived from 
equivalent width measurements of the \ion{Ca}{ii} K line.
In Table \ref{AnalysisResults} we have adopted the more conservative $\sigma_{[\rm Ca/H]}=0.4$.

However, we obtain more precise Ca abundances using SYNTHE, which 
allows us to fit the entire spectrum in more detail. In practice we
compare the observations with models computed for an array of abundances
in steps of 0.05\,dex.
 By varying the values of effective temperatures and surface gravities 
taking into account the uncertainties
previously discussed for the SDSS analysis, we derive 
uncertainties of 0.2\,dex in the SYNTHE calcium abundances. 
The final metallicity uncertainties in Section \ref{AnalysisResults} are derived from 
$\left[{\rm Fe/H}\right]=\left[{\rm Ca/H}\right]-\left[{\rm \alpha/Fe}\right]$, assuming
$\left[{\rm \alpha/Fe}\right]=0.4$

\subsection{Carbon abundance}\label{carbon}

As discussed in the Introduction, CEMP stars are specially frequent among the most metal-poor stars. 
It has been 
proposed that there is  a bimodality of carbon abundances in CEMP stars 
in the [Fe/H]$<-3.5$ regime \citep{spi13,boni15,alle15,han15III}.

A population of stars shows higher carbon abundances, A(C)$\sim8.25$ than the other, 
A(C)$\sim6.5$.
Some CEMP stars exhibit high abundances of slow neutron-capture elements 
and are defined by  $\left[{\rm Ba/Fe}\right]>+1.0$
(CEMP-s stars). On the other hand, CEMP-no stars do not show 
such high abundances of s-process elements and $\left[{\rm Ba/Fe}\right]<+0.0$ \citep{bee05II}.

Using SDSS spectra we are able to measure the carbon abundance only in few cases, when the G band is clean
enough and  the S/N ratio is high enough.

\begin{figure*}
\begin{center}
{\includegraphics[width=200 mm, angle =180]{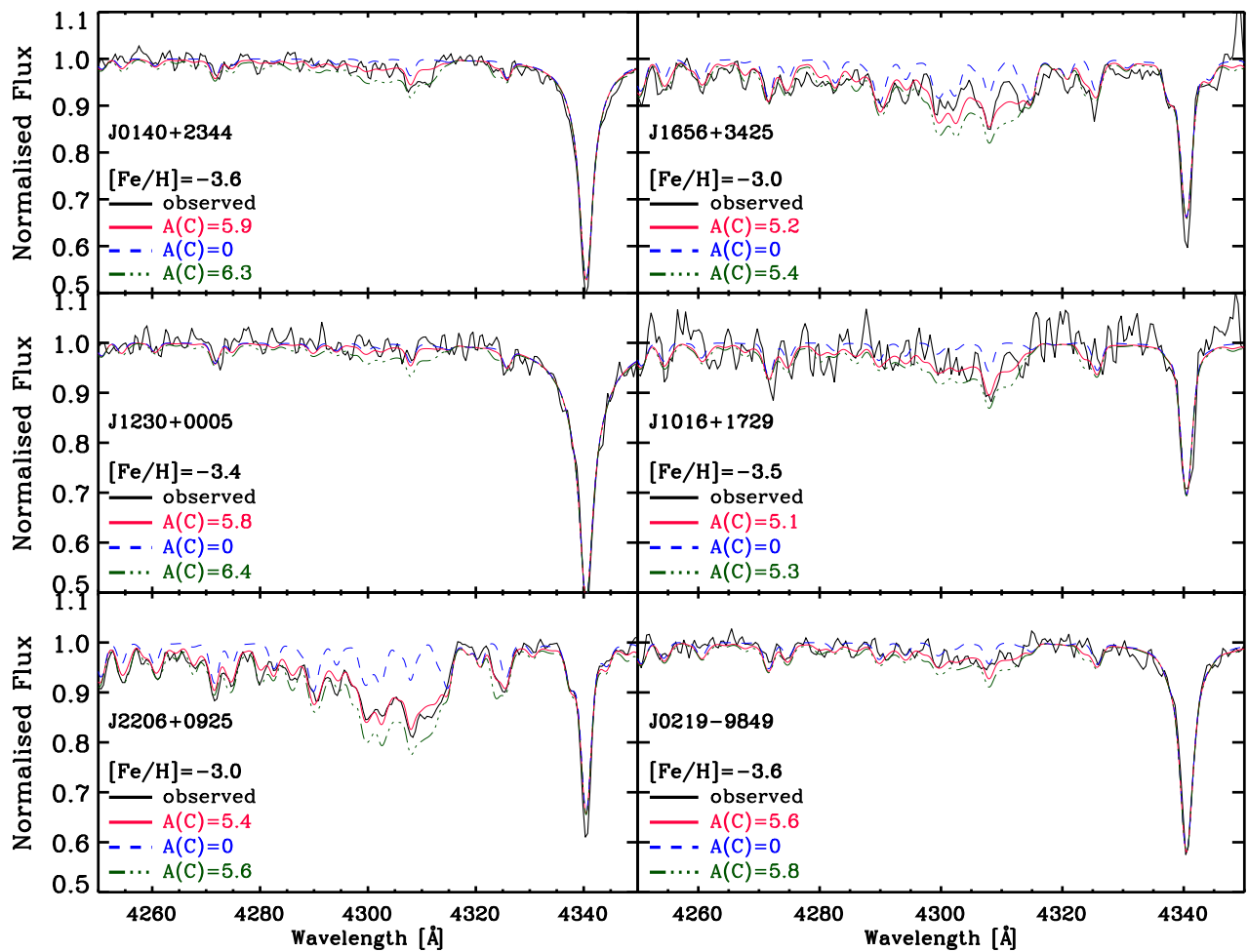}}
\end{center}
\caption{ Six stars reported in this work. 
The black line shows the observed spectrum, the red one is the best  fitting model calculated with SYNTHE, 
the blue-dashed curve corresponds to the case the stars are not enhanced in carbon
relative to solar C/Fe abundance ratios, and the green dotted line corresponds to 
upper limits. 
In four cases we are able to give a value for carbon abundance J0140+2344, 
J1016+1729, J1656+3425 and J2206+0925 and upper 
limits are given for the other two stars, J0219-9849, J1230+0005.}
\label{cempanel}
\end{figure*}

For the stars in our sample, we can use the G band at $\sim$ 4300\,\AA  to derive
carbon abundances \citep{lam84} thanks to the higher ISIS spectra resolving power and, in several cases,
  higher S/N. We manually fit the spectra with SYNTHE,   the same process followed for deriving calcium abundances.
  Our results are included in Table~\ref{AnalysisResults}.
Fig.~\ref{cempanel} shows all the cases for which we detect the G band, but only upper limits are provided in some 
cases due to the weakness of the G band in those stars.
The lower effective temperature the easier the carbon detection.

 In the bottom panel of Fig. 6 we display the [C/Fe] ratios versus the metallicity [Fe/H] 
of the stars in our sample, together with the stars from the literature.
Three of stars analysed in this work, J101600+172901, J165618+342523, J220646$-$092545, do not give extra information about the nature 
of CEMP because all are at about the
level of [C/Fe]$\sim+0.45$, and are ``normal`` metal-poor turn-off stars 
\citep{beha10I,mas10,cohe13}.
In the case of  J021958$-$084955 and J123055+000547 we are only able to give an upper 
limit to the carbon abundances. Probably this two stars are also ``normal`` metal-poor turnoff stars.
Finally J014036+234458 is the only star with $\left[{\rm C/Fe}\right]>+1.0$ and the most metal-poor 
star of this sub-sample, according to the statistics mentioned in Section \ref{Intro}.

The ISIS medium-resolution data are only in few cases able to detect the G band 
absorption. We have estimated
the uncertainties from the difference between the best fit model and 
visually identified upper limits.  This was only possible
in four cases, given in Table \ref{AnalysisResults}.

\section{Comparison with previous work}\label{compa}

Five of our targets have been observed and analyzed in 
recent independent studies. 
As explained in \S \ref{TargetSelection} and \S \ref{wht_analysis}, we derive 
the calcium abundance from the resonance Ca II K line and assume [$\alpha$/Fe]$=+0.4$ 
to derive a metallicity. Below we compare our results with those in the literature.

\subsection{SDSS J014036+234458}\label{wht_w2044}

\begin{figure*}
\thispagestyle{empty}
\begin{center}
{\includegraphics[width=165 mm]{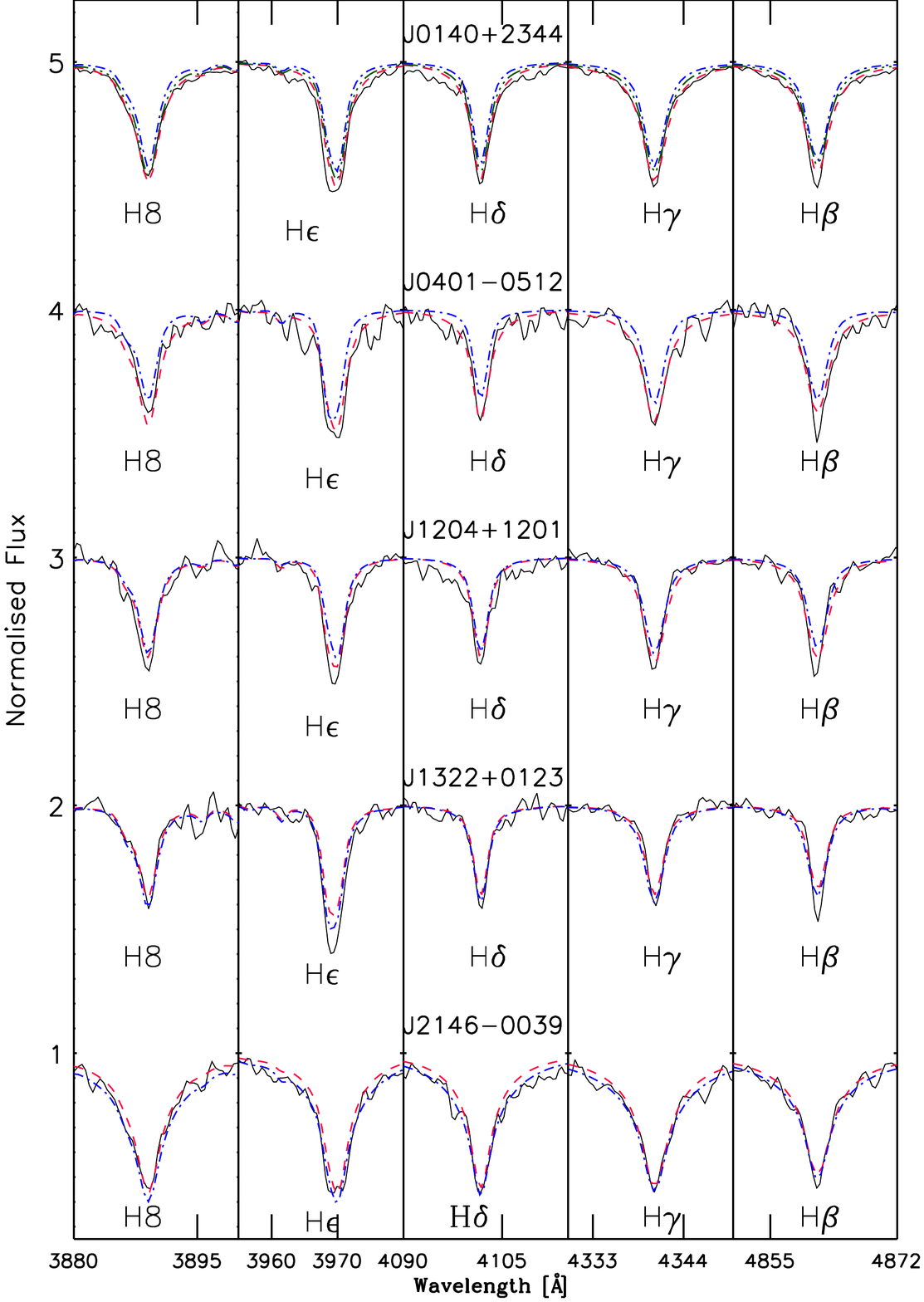}}
\end{center}
\caption{
ISIS spectrum of the stars J0140+2344, J0401-0512,J1204+1201, J1322+0123 and J2146-0039
(black solid lines) and best fit obtained 
in this work (red dashed line) performed with ASSET with the set of parameters  $\langle T_{\rm eff},\logg,\left[{\rm Fe/H}\right]\rangle$,
[6090,4.7,-3.6], [5850,5.0,-3.6], [5850,3.9,-3.7], [5250,2.0,-3.7], [6470,4.9,-3.6], respectively.
Two additional synthetic spectra are depicted for J0140+2344: one (green triple-dotted-dashed-line) using 
$T_{\rm eff}=$ 5848\,K, $\logg=4.0$, [Fe/H]$=-3.83$ \citep{caff13I} and other spectrum (blue triple-dotted-dashed-line) $T_{\rm eff}=$ 5703\,K, $\logg=4.0$, [Fe/H]$=-4.00$ \citep{yong13II}.
One spectrum for the other objects (blue dotted-dashed line) with values: [5500,4.0,-3.6] for 
J0401-0512 \citep{caff13I}, [5467,3.2,-4.34] for J1204+1201 \citep{pla15}, [5466,3.12,-3.32] for J1322+0123 \citep{pla15} and
[6475,4.0,-3.14] for J2146-0039  \citep{caff13I}.}
\label{w3157}
\end{figure*}

The analysis of the SDSS and ISIS spectra of J0140+2344 
leads to  
 $T_{\rm eff}=6090\pm200$\,K, $\log g=4.7\pm0.3$ and [Fe/H]$=-3.6\pm0.2$.
 The continuum slope fit suggests (See Fig. \ref{compared} the general uncertainty
 adopted in \ref{TargetSelection} have to increase to 200\,K.
However \citet{yong13II}
uses a high resolution spectrum obtained with HIRES at the Keck-I telescope to derive  $T_{\rm eff}=5703\pm85$\,K
and [Fe/H]$\simeq -4.0$.
On the other hand, \citet{caff13I} using X-SHOOTER at VLT spectra derive $T_{\rm eff}=5848$\,K, and [Fe/H]$=-3.83$ assuming $\logg=4.0$.
The differences among the three[Fe/H] values are most likely related to the 
different adopted temperatures.
In Fig.~\ref{w3157} we display the ISIS spectrum of this star, together with three synthetic spectra.
The best fitting of the Balmer lines appears to correspond to our $T_{\rm eff}$ 
value. 
While the SYNTHE code use \citet{ali66} theory  for self-broadening of Balmer lines in order to derive effective 
temperature \citet{yong13II, caff13I} use the \citet{ber00c,ber00b} theory. 
In principle, \citet{sbo10} showed about +350 \,K disagreement at the level of 5800 \,K and +200\,K at 6100\,K,
 between effective temperatures derived from H$\alpha$. Using both theories, providing the
the \citet{ali66} theory hotter temperature scale.
 However we note here that our T$_{\rm eff}$ has been derived with {\tt FER\reflectbox{R}E}, 
using a grid of model synthetic spectra computed with the ASSET code \citep{koe08},
which also uses the Barklem theory for Balmer lines but also the slope of the continuum.
In Fig. \ref{w3157} we plot synthetic spectra for different set of parameters using ASSET.
 Our C abundance, $A(\rm C)=5.9\pm 0.4$, which is compatible 
with the reference value from the literature, $A(\rm C)=5.56\pm0.04$,
was derived assuming the star is a dwarf \citep{yong13II}. Moreover, \cite{caff13I} only give an upper limit 
for carbon abundance of $A(\rm C)\leq6.0$.

\begin{table}
\setlength{\tabcolsep}{4pt}
\label{w2044}
\caption{ J0140+2344 atmospheric parameters.}
\centering
\begin{tabular}{llllll}
\hline\hline
Author& $T_{\rm eff}$ &$\left[{\rm Fe/H}\right]$ & $\log g$ &$\left[{\rm C/Fe}\right]$\\
\hline
 \citet{yong13II}&5703&-4.0&4.68 &+1.13 \\
\citet{caff13I} &5848&-3.83&4.0 & $\leq1.4$ \\
This work &6090&-3.6&4.77 & +1.07 \\

\hline
\end{tabular}
\end{table}

\subsection{SDSS J040114$-$051259}

 The metallicity derived by \citet{caff13I}, $\left[{\rm Fe/H}\right]=-3.62$, is the same as that derived in this work.
Nevertheless, the derived effective temperature, $T_{\rm eff}=$ 5500\,K, is significantly different from our value, 
 $T_{\rm eff}=$ 5850\,K.
In Fig.~\ref{w3157} we depict the synthetic spectra with both sets of parameters and the best-fit 
calcium abundance. 
 The synthetic spectrum, even though computed with ASSET, using our stellar parameters seems to well reproduce better the ISIS spectrum of this star (see Fig \ref{w3157}).

\subsection{SDSS J120441+120111}\label{3214}

The main stellar parameters derived from the analysis of SDSS spectra by 
\citet{pla15}, $T_{\rm eff}=$ 5894\,K, 
$\log g=2.66$, [Fe/H]$=-3.41$ are in fair agreement with our own results: $T_{\rm eff}=$ 5850\,K, 
$\log g=3.9$, and [Fe/H]$=-3.7$. However the authors proposed a different set of parameters based on a high-resolution spectrum 
obtained using MIKE spectrograph at the 6.5m Magellan telescope,  
$T_{\rm eff}=$ 5467\,K, $\log g=3.20$  and [Fe/H]$=-4.34$.
 This metallicity difference of about 0.6\,dex is easily explained by the 400\,K difference.

 \citet{pla15} provide the equivalent widths for 22 iron lines, the Ca K line and the \ion{Ca}{i} transition at 4226 \AA.
Assuming our atmospheric parameters we have derived the following values using \emph{Abfind} routine with MOOG:
$\left[{\rm \ion{Fe}{i}/H}\right]=-3.9$, $\left[{\rm \ion{Ca}{i}/H}\right]=-3.7$ and
$\left[{\rm \ion{Ca}{iI}/H}\right]=-3.8$ while, acording to our SYNTHE analysis, 
we derive $\left[{\rm \ion{Ca}{iI}/H}\right]=-3.3$ for the resonance calcium K line.
As described in Section 2, our our $T_{\rm eff}$ determination is based on simultaneous 
fitting of the stellar continuum and Balmer lines and therefore
we consider it more reliable than the high resolution analysis which
is based on excitation equilibria of FeI lines and after corrected procedure explained in 
 \citet{pla15}, and references thererin.

\subsection{SDSS J132250+012343}

Following a similar process as in the case of J1204+1201, \citet{pla15} derived 
 $T_{\rm eff}=$ 5466 K, $\log g=3.12$, and [Fe/H]$=-3.32$ 
from its SDSS spectrum. 
This is somewhat different from our results based on the same data: 
$T_{\rm eff}=$ 5250\,K, $\logg=2.0$, $\left[{\rm Fe/H}\right]=-3.7$.
The same authors, using a high resolution spectrum, derived a slightly different set of parameters,  
$T_{\rm eff}=$ 5008\,K, $\logg=1.95$, $\left[{\rm Fe/H}\right]=-3.64$ which are compatible with our own results.
 The equivalent widths measured by \citet{pla15} with our
stellar parameters lead to $\left[{\rm \ion{Ca}{i}/H}\right]=-3.2$ 
and $\left[{\rm \ion{Fe}{i}/H}\right]=-3.3$.

\subsection{SDSS J214633$-$003910}

The SDSS spectrum of J2146-0039  has a poor quality in terms of low signal-to-noise ratio. 
However, our pipeline derives $T_{\rm eff}=$ 6470\,K, 
in good agreement with the value reported by \citet{caff13I}, 
who find $T_{\rm eff}=$ 6475 K. 
In addition, both analyses suggest that J2146-0039 is a dwarf.

The metallicity proposed by \citet{caff13I}, derived using a 
high-resolution spectrum, 
is [Fe/H]$=-3.14$, while our prefered value is 0.4\,dex lower, [Fe/H]$=-3.6$.
We are able to resolve a significant contribution to the observed feature 
from interstellar calcium 
at 3933\,\AA. A more detailed analysis in order to resolve this discrepancy is needed.

\section{Analysis of HET spectra with SYNTHE}

High-resolution ($R\sim 15,000$) spectroscopic observations of the stars
 J0140+2344
and J0219$-$0849
were carried out 
in order to determine the abundances of Fe, Mg, and Sr in these stars. 
The effective temperatures and surface gravities from our analysis of 
their SDSS spectra were adopted.  For both stars, the signal-to-noise  
of the spectra from the HRS 
red chip (4800\,\AA- 5400\,\AA) is higher than those from the blue one (4000\,\AA- 4700\,\AA),
with only a few elements that could be analyzed. 
We find that our results from HRS/HET observations are 
compatible with those from ISIS.
We include the derived chemical abundances in Table \ref{het}.

\subsection{J014036+234458}\label{het_w2044}

\citet{yong13II} provide complete information 
about the chemical abundances 
derived using the stellar parameters discussed in \S \ref{wht_w2044}. 
According to our results, the discussion in this section 
assumes J0140+2344 is  a  
dwarf star ($\logg=4.7$).
The only \ion{Fe}{i} line we are able to measure at 4328\,\AA gives us the iron abundance A(\ion{Fe}{i})=3.8 (see table \ref{het})
or [Fe/H]=-3.6 in perfect agreement with the metallicity derived in Section \ref{wht_analysis}.
We propose a magnesium abundance A(\ion{Mg}{i})$=4.4\pm0.2$ while \citet{yong13II} 
derived $3.86\pm0.06$. 
There is a gap of about 0.5\,dex between these two values, which  could be
explained by 
the difference  effective temperatures adopted in the two analyses. 
For \ion{Ca}{i} we can only  
set an upper limit A(\ion{Ca}{I})$<3.5$, while \citet{yong13II} 
determined A(Ca)$=2.53\pm0.06$.
A similar situation is found for Sr, with an abundance of A(\ion{Sr}{II})$=0.01\pm0.06$ 
obtained by \citet{yong13II}, 
compared to our upper limit of A(\ion{Sr}{II})$<0.5$.
It seems the 0.3\,dex gap in metallicity and various elemental abundances 
are closely linked to the difference of 300~K in effective temperature.
\subsection{J021958$-$084955}\label{het_w3122}
\begin{figure*}
\begin{center}
{\includegraphics[width=90 mm, angle =180]{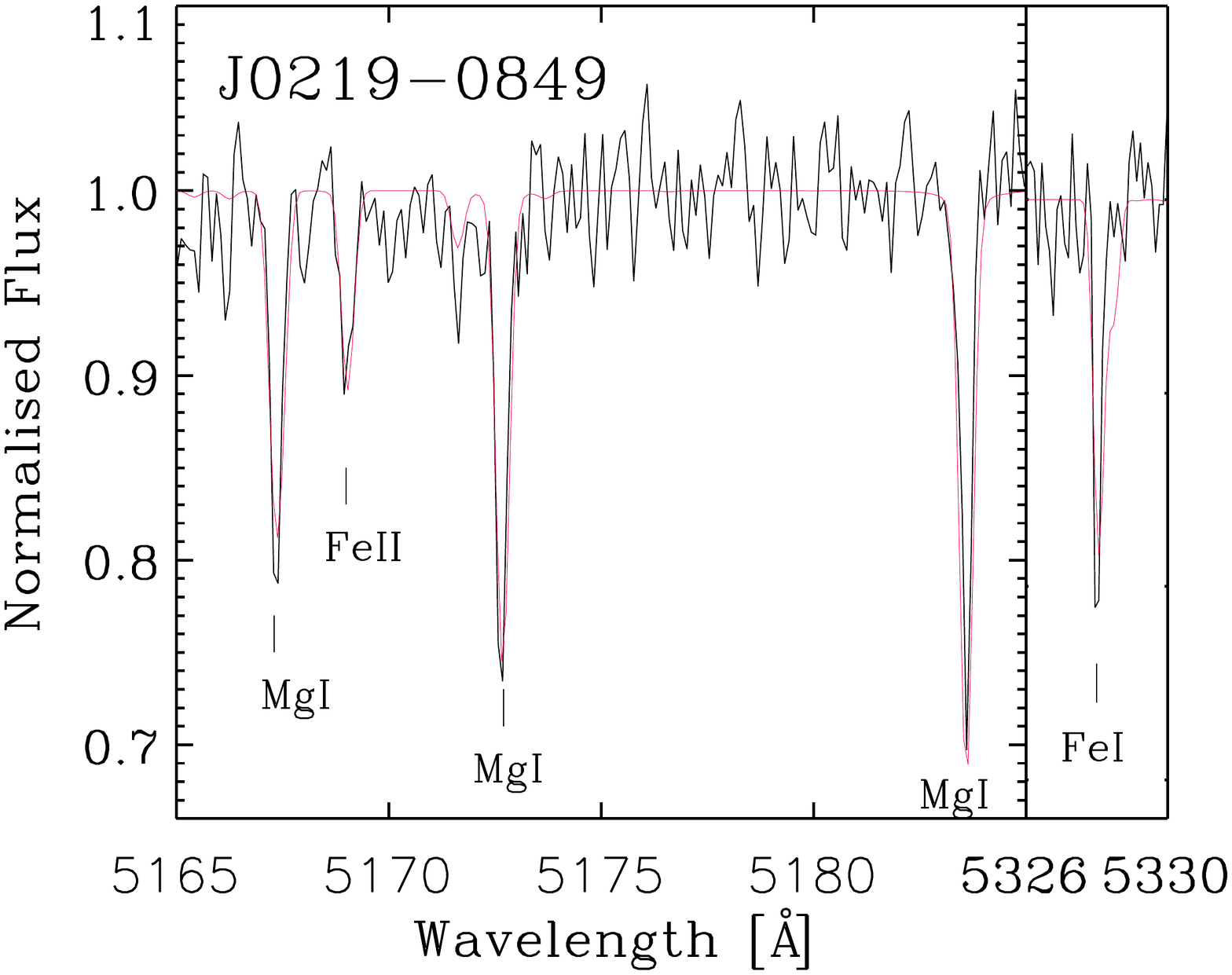}}
{\includegraphics[width=90 mm, angle =180]{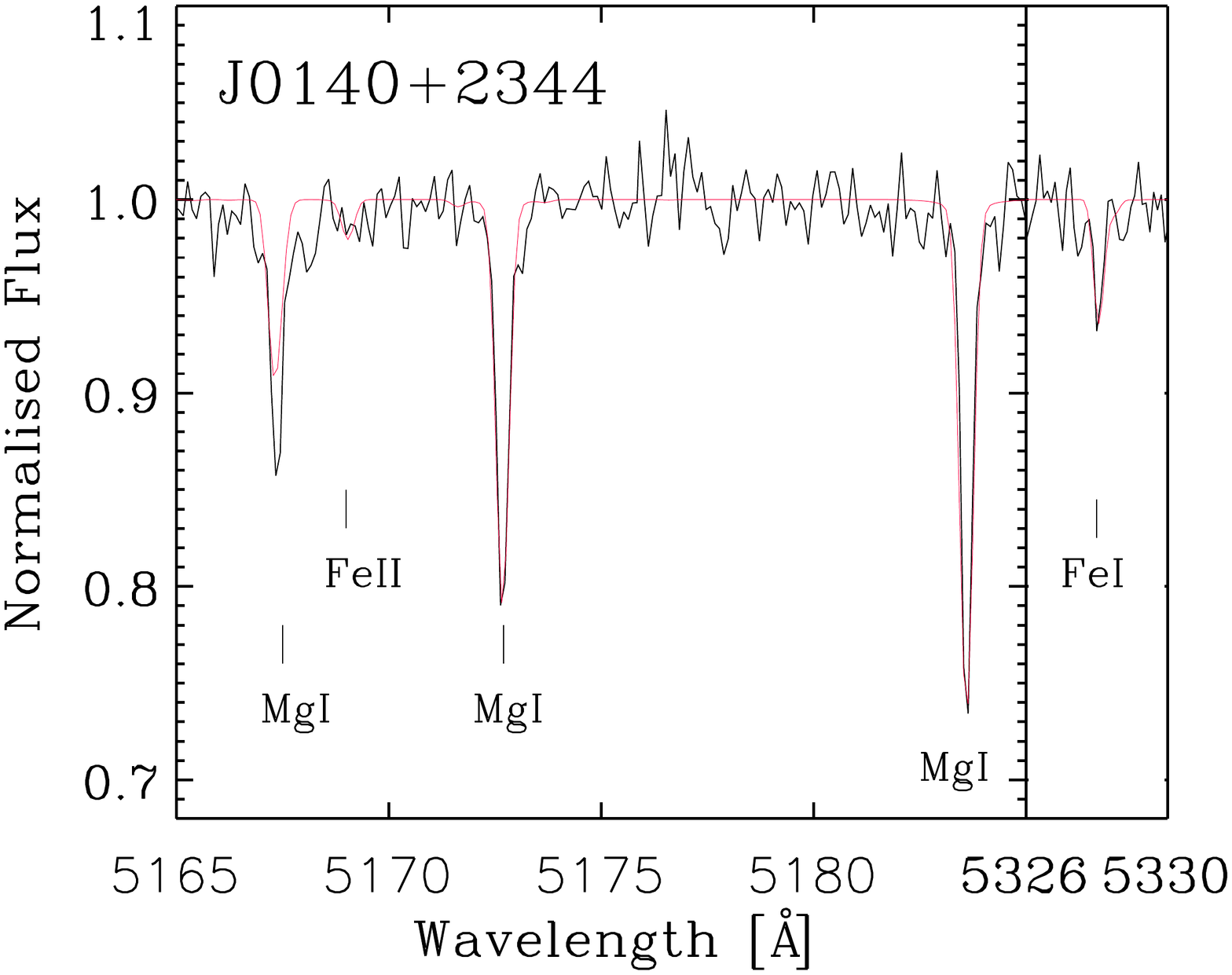}}
\end{center}
\caption  {Observed HRS spectra (black line) and best-fit synthetic spectra (red lines) of
the  \ion{Mg}{i}b triplet region in the stars J0219-0849 (left panel) and J0140+2344 (right panel).}
\label{magnesium}
\end{figure*}

The red part of the spectrum allows us to measure the magnesium abundance and 
a single iron line (5325\,\AA). Upper limits for other chemical abundances 
are given 
in Table~\ref{het}. 
In Fig.~\ref{magnesium} we depict the observed HRS spectra
of the stars J0219-0849 and J0140+2344, together with 
synthetic spectra for our best-fitting parameters.
The uncertainties involved in  high resolution spectroscopic
analysis have been discussed 
by \citet{alle08}. 
Our error bars are included in Table~\ref{het}, where only  
upper limits are given for some elements. 

\begin{table}
\setlength{\tabcolsep}{4pt}

\caption{ HET abundances.\label{het}}
\centering
\begin{tabular}{llllll}
\hline\hline
Object& A(\ion{Mg}{i})& A(\ion{Ca}{i})& A(\ion{Sr}{II})& A(\ion{Fe}{i})& A(\ion{Fe}{ii})\\
\hline
J0140+2344 &$4.4\pm0.2$ &$<3.5$&$<0.5$ &$3.8\pm0.2$ & $ <3.3$\\
J0219$-$0849 &$4.2\pm0.2$&$<3.2$&$<0.7$ & $4.0\pm0.2$ &  $ 3.9\pm0.2$\\

\hline
\end{tabular}
\end{table}

\section{Discussion and conclusions}
\label{Conclusions}
 
We carry out a combined analysis of 
SDSS  
and ISIS/WHT spectroscopy to identify several 
extremely low metallicity stars. The candidates are selected after
analysis of the SDSS data, and followed up with high quality ISIS/WHT and HRS/HET observations.
From the comparison of the metallicities we inferred from SDSS spectra with those from our analysis of 
ISIS/WHT data, HRS/HET data, and the literature when available, 
we conclude that our selection based on SDSS spectra  
is highly reliable.
In Fig. \ref{w3157} we plot the carbon abundance for several CEMPS 
with [Fe/H]$<-4.0$, and several more in the  [Fe/H] $<-3.0$
regime.
Two of our objects, J\,1656+3425 and J\,2206-0925, 
are giant stars and it is possible  that 
their carbon abundance can be somewhat 
affected by mixing in deep layers \citep{spi06}.

The two main sub-classes  of the CEMP stars (the CEMP-no and CEMP-s stars)
 are discussed in Section \ref{carbon}. 
\citet{han15III} suggest that CEMP-s stars are generally the product of mass transfer from a asymptotic giant-branch 
(AGB) star companion.
This mass transfer could be the origin of the carbon enrichment, which would
  explain why most of the CEMP-s 
stars exhibit very high carbon abundances \citep{boni15}.
Over 90\% of the stars located in the band with very high carbon enhancements
are CEMP-s or CEMP-rs. The source of the s-process elements in CEMP-s stars are not well explained yet. 
Nevertheless, the carbon in CEMP-no stars was provided by their natal molecular clouds \citep{sta14}. 
In fact, CEMP-no stars could be part of  binary systems, 
but their enhanced carbon does not have an origin in a mass-transfer
process. The aforementioned authors refer to these objects as 
bona-fide fossil records because the abundances in these objects
reflect those in the interstellar medium where they were formed \citep{boni15, han15II}.

Several questions related to the formation and composition of EMP, UMP and HMPs still 
remain unresolved. The fact that SDSS J102915+172927 \citep{caff11} is the only 
known UMP unevolved star with [C/Fe]<+1 is not yet understood. 
On the other hand,  faint supernova models (see e.g. \citet{tom14})
or models of \emph{spinstars}, massive rotating metal-poor stars, see e.g. \cite{mae15II}, seem to explain the main features of the abundance patterns of CEMP,
CEMP-no stars, but surely still more identification of stars at extremely low metallicities 
are need to better understand the formation of these stars in the early Universe.
This paper describes our observational program.
Additional observations are taking place and  will be used to try to answer some of the previous
questions.

\begin{acknowledgements}

DA acknowledges the Spanish Ministry of Economy and Competitiveness (MINECO) for the financial support
received in the form of a Severo-Ochoa PhD fellowship, within the Severo-Ochoa International PhD Program.
DA, CAP, JIGH, RR also acknowledge the Spanish ministry project MINECO AYA2014-56359-P. 
JIGH acknowledges financial support from the Spanish Ministry of Economy and Competitiveness (MINECO)
under the 2013 Ram\'on y Cajal program MINECO RYC-2013-14875. 
DLL acknowledges the support of the Robert A. Welch
Foundation through grant F-634. EFA acknowledges support from DGAPA-UNAM postdoctoral fellowships.

This paper is based on observations made with the William Herschel
Telescope, operated by the Isaac Newton Group at the Observatorio del Roque de los Muchachos, La Palma,
Spain, of the Instituto de Astrof{\'i}sica de Canarias. We thank ING staff members for very efficiently 
during the four observing runs in visitor mode.  \\
The Hobby-Eberly Telescope (HET) is a joint project of the University of Texas at Austin, the Pennsylvania State University, Stanford University, Ludwig-Maximilians-Universit\"at M\"unchen, and Georg-August-Universit\"at G\"ottingen. The HET is named in honor of its principal benefactors, William P. Hobby and Robert E. Eberly.

\end{acknowledgements}


\bibliography{biblio}

\begin{appendix}

 \section{Tables}
We present a table with observing runs details.

 \begin{table*}
\renewcommand{\tabcolsep}{5pt}
\centering

\caption{
The ISIS/WHT and HRS/HET Observing log.\label{observations}}
\begin{center}
\begin{tabular}{lcrccc}
\hline
Star                 & $g$ & $N_{\rm exp}$ x $t_{\rm exp}$ & observing run &literature& S/N   \\
                     & mag &   & && at 4500\,\AA     \\
\hline\hline
ISIS/WHT\\
\hline
SDSS\,J010505+461521   & 19.3 & 13x1200\,s & I && 38    \\
SDSS\,J014036+234458   & 15.8 & 2x1200\,s  & I&1,2& 90     \\
SDSS\,J021958$-$084955 & 16.4 & 4x1200\,s  & I&& 80       \\
SDSS\,J040114$-$051259 & 18.6 & 5x1800\,s  & IV&2& 37        \\
SDSS\,J044655+113741   & 18.6 & 4x1200\,s  & I& &22       \\
SDSS\,J063055+255243   & 18.1 & 4x1800\,s  & IV&& 40       \\
SDSS\,J075818+653906   & 17.7 & 6x1800\,s  & IV&& 60       \\
SDSS\,J101600+172901   & 16.6 & 1x1200\,s  & II&& 33       \\
SDSS\,J120441+120111   & 16.4 & 2x1800\,s  & II&4& 50     \\
SDSS\,J123055+000547   & 14.8 & 1x600\,s    & II&& 58    \\
SDSS\,J132250+012343   & 16.3 & 1x2400\,s  & III&4& 60       \\
SDSS\,J134338+484426   & 12.5 & 2x1800\,s  & III&& 487     \\
SDSS\,J164234+443004   & 17.8 & 5x1800\,s  & IV&& 39       \\
SDSS\,J165618+342523   & 15.7 & 2x1800\,s  & II& &70    \\
SDSS\,J183455+421328   & 19.1 & 4x1800\,s  & I& &24      \\
SDSS\,J214633$-$003910 & 18.1 & 5x1200\,s  & I&2& 40       \\
SDSS\,J220646$-$092545 & 15.4 & 2x1200\,s  & I&& 110    \\
\hline
2MASS\,J204523$-$284235 & 14.0  &1x1200\,s & II&1  & 118  \\
\hline
\hline
HRS/HET\\
\hline
SDSS\,J014036+234458 & 15.8 & 4x1500\,s& 11/12 dec 2012&1,2&80 at 5500\,\AA\\
SDSS\,J021958$-$084955 & 16.4 & 8x3400\,s&6 dec 2012 /13 feb 2013&&52 at 5500\,\AA \\
\end{tabular}
\end{center}
\begin{tablenotes}
 \item   $^{1}$\citet{yong13II}; $^{2}$\citet{caff13I};$^{3}$\citet{caff11} $^{4}$\citet{pla15};
 $^{5}$\citet{fre05}; $^{6}$\citet{fre06}
\end{tablenotes}
\end{table*}

\end{appendix}

\end{document}